\documentclass[final,5p,times]{elsarticle}
\usepackage{lineno}
\usepackage{lineno}
\usepackage[breaklinks,hidelinks]{hyperref}\hypersetup{colorlinks,allcolors=black}
\usepackage{graphicx}
\usepackage{amssymb}
\usepackage{siunitx}
\DeclareSIUnit\clight{\text{\ensuremath{c}}}
\DeclareSIUnit[number-unit-product = ]\percent{\char`\%}
\usepackage{pdfpages}
\usepackage{subcaption}
\usepackage{float}
\usepackage{amsmath}
\usepackage{color}
\usepackage{ulem}
\usepackage{footnote}
\usepackage{textcomp}
\usepackage{xcolor}
\usepackage{xspace} 
\usepackage{booktabs}
\usepackage{multirow}
\usepackage{todonotes}
\usepackage{wasysym}
\usepackage{tabularx}
\usepackage{mathtools}
\usepackage{verbatim}
\usepackage{breakurl}

\newcommand{\secref}[1]{Section~\ref{#1}}

\newcommand{\tabref}[1]{Table~\ref{#1}}

\newcommand{\NeCOtwo}{Ne-CO$_2$ (90-10)\xspace}

\newcommand{\ArCOtwo}{Ar-CO$_2$ (90-10)\xspace}
\newcommand{\ArCOtwoThirty}{Ar-CO$_2$ (70-30)\xspace}

\newcommand{\qcrit}{\ensuremath{Q_\mathrm{crit}}\xspace}

\newcommand*{\eg}{e.\,g.\@\xspace}

\makeatletter\newcommand*{\etc}{etc\@ifnextchar.{}{.\@}}\makeatother

\journal{NIM A}
\title{Discharge mitigation methods in MPGD-based detectors}
\begin{document}
\begin{frontmatter}

\author[a,b]{P.~Gasik\corref{cor1}}\ead{p.gasik@gsi.de}


\address[a]{GSI Helmholtzzentrum f\"{u}r Schwerionenforschung GmbH (GSI), Darmstadt, Germany}
\address[b] {Facility for Antiproton and Ion Research in Europe GmbH (FAIR), Darmstadt, Germany}

\cortext[cor1]{Corresponding author}
\begin{abstract}

This work reviews various methods used to minimize the probability of discharge occurrence in MPGDs or to mitigate the effect of
spark discharges on the detector system. These include techniques that can reduce the probability that the fundamental discharge
limits of MPGD structures are reached during detector operation, methods based on
the HV scheme optimisation, or the implementation of resistive electrodes to enhance local discharge quenching capabilities in an amplification structure. In addition, design optimization and quality control methods to ensure stable, long-term operation
of an MPGD-based detector, are considered.

\end{abstract}
\end{frontmatter}

\section{Introduction}
\label{sec:intro}

The requirements of a new generation of experiments are driving factors for developing new detectors. Novel devices must handle steadily increasing particle rates, providing high sensitivity in a wide dynamic range for detecting minimum ionizing particles and highly ionizing nuclear fragments. Outstanding particle identification capabilities of time projection chambers (TPC), or large-area high-rate trackers are still the domain of the gaseous detector technology providing cost-effective solutions characterized by great performance figures. Among many new innovative technologies, Micro-Pattern Gaseous Detectors (MPGD) have become widely used in nuclear physics experiments and are foreseen at future facilities.

Despite many years of experience in the production and successful operation of MPGDs, the high demands of modern experiments, which, in particular, concern the substantial increase in active detector area, require further developments and improvements. Constantly increasing luminosity demands operation in a harsh, high-rate environment posing stringent requirements on the performance of gaseous detectors including, depending on the use case, high gain and high efficiency, spatial, time and energy resolutions, or ion backflow capabilities. The key parameters for long-term operation are radiation hardness, aging resistance, and stability against electrical discharges. The latter in particular poses a threat to the integrity of the detector and its readout electronics, as they may cause irreversible damage to the amplification structure, ranging from enhanced leakage currents to permanent electric short circuits that render the detector effectively blind and lead to a loss in its acceptance. %
It is, therefore, of the highest importance to further study and optimize this particular aspect of detector operation. By quantitatively understanding underlying mechanisms, the probability of the development of discharge events can be substantially reduced allowing for a stable operation under extreme conditions. In the following, we present an overview of well-established mitigation strategies for the safe operation of MPGD-based detectors.

\section{Discharges in Micro-Pattern Gaseous Detectors}
\label{sec:mpgddisch}
The main mechanism of discharge formation in MPGD structures (see recent review in \cite{sauli2020book}) is the streamer mechanism~\cite{Raether1939, loeb1939fundamental, doi:10.1063/1.1707290, PhysRev.57.722}, and the development of a spark discharge. Indeed, it is observed that a streamer develops in a micro-pattern detector when the total number of charges in an avalanche reaches a critical charge limit \qcrit of \SIrange[range-units=single]{e6}{e8}{e}~\cite{peskov958725, bressan1999high, fonte2010physics, sauli_amsterdam}. However, it should be noted, that in none of these references (and many others, not mentioned here) a universal \qcrit value can be specified for a given MPGD structure. Reported critical charge values and discharge probabilities depend not only on the geometry of the structure, or the size of its amplification gap, but also on the type of radiation source (number of primary electrons)~\cite{peskov958725} and its inclination~\cite{BACHMANN2002294}, or the exact composition of the gas mixture~\cite{bressan1999high, DELBART2002205}. In the latter, a clear correlation between the discharge probability and the average atomic number $\langle Z\rangle$ of the gas mixture is observed pointing to low-$Z$ gases being more stable concerning spark discharge development. It is therefore clear, that the local charge densities approaching single amplification cells (\eg a GEM hole) play a key role in the discharge development and the long-term stability of an MPGD structure. 

The latter has been recently studied measuring discharge stability of GEMs, THGEMs, and Micromegas in  Ne- and Ar-based gas mixtures~\cite{GASIK2017116, Gasik:2022vft, Gasik_2023}. In all cases, a clear gas dependence is observed pointing to low-$Z$ mixtures as more stable ones. Surprisingly, it is also observed that increasing the content of a quencher gas does not always improve the stability which, again, can be explained by the average atomic number of the mixture and the local charge densities obtained therein~\cite{Gasik:2022vft}. A simulation framework was developed that allows estimating critical charge limits from the comparison to the measured data. The values of the \qcrit parameter have been extracted for all the gas mixtures used in the measurements with GEMs and THGEMs and are summarised in \tabref{tab:Qcrit}.
\begin{table}[tbp]
    \centering 
    \caption{Critical charge values extracted from the simulation fits to the experimental data for GEMs~\cite{GASIK2017116} and THGEMs~\cite{Gasik:2022vft}.}
\begin{tabular}{lcc}
\toprule
  \multirow{2}{*}{Gas} & \qcrit THGEM & \qcrit GEM \\[.5ex]
  & $[\times10^6\,e]$ & $[\times10^6\,e]$ \\
  \midrule
   \NeCOtwo & 7.1 $\pm$ 2.2 & 7.3 $\pm$ 0.9 \\[1ex]
   \ArCOtwo & 4.3 $\pm$ 1.5 & 4.7 $\pm$ 0.6 \\[1ex]
   \ArCOtwoThirty & 2.5 $\pm$ 0.9 & --\\
   \bottomrule
   \label{tab:Qcrit}
\end{tabular}
\end{table}
The results confirm the observation of the \qcrit gas dependency. Moreover, the values obtained for both amplification structures nicely agree with each other, despite the clear geometrical differences and different electric field configurations inside GEM and THGEM holes. The primary charge limits shall be therefore considered per single hole and not normalized to the hole volume. This would support the hypothesis that the effective volume of streamer formation is similar in both cases~\cite{Gasik:2022vft}. 

Both GEM and THGEM measurements support the conclusion that the primary charge density, arriving at the single amplification cell, is a key factor influencing the stability of a GEM-like structure against a spark discharge. In the recently launched Micromegas studies, we show that similar conclusions can be drawn in the case of mesh-based structures, and the mesh cells can be considered as independent amplification units~\cite{Gasik_2023}.

\section{Discharge mitigation methods}
\label{sec:methods}

There is no single way to minimize the probability of discharge occurrence or to mitigate the effect of spark discharges on the detector system. Depending on the type of detector, its purpose, design details, and experimental conditions, there may be different ways to avoid or mitigate electrical discharges. There are, however, several groups of methods one can define to describe different approaches. While the fundamental, physics-driven stability limits of an MPGD cannot be easily changed, the system optimization and quality control are continuously being improved and with more than 25 years of experience become more and more effective. Methods presented in the following are based on well-established principles and have been further worked out in the course of the recent large-scale detector developments~\cite{Adolfsson_2021, ABBAS2022166716, ALLARD2022166143}.

\subsection{Choice of gas and operational conditions}
\label{sec:opti:gas}


A gas mixture is a basic ingredient of any gaseous detector. Its final choice is based on thorough consideration of many requirements, drift and diffusion coefficients, amplification, attachment, or aging properties being only a few of them. Together with the optimized HV settings, they define the detector performance such as gain, position and energy resolutions, ion backflow, \etc. The necessity of stable operation in a high-radiation environment requires often contradictory sets of parameters to be considered, which may compromise the detector's performance. On the other hand, long-term operation in a hermetic detector setup is a must for high-efficiency running. 

In general, gas optimization is strongly related to the intrinsic stability limits of the amplification structure and the gas itself. One can summarise their optimization concerning the spark discharge occurrence as follows.

\begin{itemize}
\itemsep0em 
    \item Choice of light gases: since a spark discharge is the most common in MPGDs, light noble gases are preferable for operation. 
    
    \item Care should be taken while choosing the quencher type and content to optimize primary charge density and electron transport properties. Note, that poor quenching capabilities of CO$_2$ imply photon and ion feedback issues, especially at high gains, which can be minimised by employing heavier quenchers (\eg iso-butane). In this case, however, primary electron densities may need to be again considered.
   
    \item Gain reduction: a trivial, however the most efficient, method to minimize the discharge probability, is to lower gain as much as allowed by the signal-to-noise ratio requirements.
  
    \item Electric field optimization above/below the MPGD structure: if allowed by the measurement requirements, the minimum diffusion coefficient and maximum drift velocity regions shall be avoided. Electron collection and extraction efficiencies shall be also considered as they directly influence the total number of charges entering a single amplification cell.
   
    \item Absolute voltage values: taking into account possible defects, residual contamination of MPGD structures, high absolute voltages, and high fields shall be avoided. In gases like neon (preferable from the spark development point of view) a glow discharge may develop at relatively low voltage values causing a constant current flow and a trip of the power supply, or damage to the detector. For this reason, high-quality production is mandatory.
    
\end{itemize} 

\subsection{MPGD stacks}
\label{sec:opti:stacks}

The most common way to enhance the gain capabilities of an MPGD-based detector is to build stacks. Among others, GEM-like structures are obvious candidates for this mechanical solution, due to their construction principles, including GEMs stretched on individual frames or THGEMs made of a thick, self-supporting PCB material. 
Structures that do not allow for easy stacking, such as Micromegas\footnote{Although double-mesh structures can be built~\cite{JEANNEAU201094, QI2020164282} their dimensions are still not sufficient for large area systems.}, or cannot be stacked being an integral part of the readout electrode (\eg MSGC~\cite{OED1988351} or WELL-like structures~\cite{Arazi_2012, Rubin_2013, Arazi_2014, Bencivenni_2015}), can be used in hybrid constructions where they play a role of the main amplification stage with a pre-amplification by GEM or a THGEM mounted above.

The aim of stacking several structures is to share the charge cloud between many individual amplification cells (\eg GEM holes) utilizing the charge diffusion and stage the gain process in subsequent detector layers. This way, individual stages can be operated at lower gains, and the probability of reaching critical charge limits in a single amplification unit can be substantially reduced. This allows obtaining much higher effective gains, although the fundamental \qcrit value for a given structure does not change.

 A stack operation requires, however, a thorough optimization of voltages, taking into account the non-trivial interplay between amplification fields, transfer fields between two MPGD layers, a drift field above the stack, and an induction field between the last amplification stage and a readout electrode. They all influence the electron and ion transport properties as well as collection and extraction efficiencies which, together with the multiplication of an individual MPGD, results in an effective gain of the entire structure (see for example~\cite{BACHMANN1999376} or more recent~\cite{Roy_2021, ratzaphd}). In addition, there are mechanical constraints to be considered while building an MPGD stack. Also in this case the reduction of discharge probability is a matter of finding a compromise between the overall performance and the long-term stability of a detector. 

\subsubsection{Multi-GEM}
\label{sec:opti:stack:gem}

In multi-GEM systems, the influence of transfer fields and charge sharing between GEM foils may seriously alter the discharge probability. The maximum gain reached in a triple GEM configuration supersedes the one of a single GEM by more than two orders of magnitude, which is a remarkable improvement in the detector performance~\cite{BACHMANN2002294}.

Conventional triple GEM systems have been operated reliably in high-rate experiments~\cite{ALTUNBAS2002177, BENCIVENNI2002493, Bozzo1462231} with the so-called \textit{standard} HV configuration, highly optimised for the minimum discharge probability. In this configuration, the gain of subsequent GEMs is decreasing towards the bottom of a stack, and thus the probability of exceeding the maximum charge limits is significantly reduced~\cite{BACHMANN2002294}. Further improvement of stability can be achieved by stacking more than three GEMs. For example, quintuple GEM stacks, considered for RICH detectors at the future electron-ion collider (eIC), are reported to operate reliably in the beam at gains exceeding $10^5$~\cite{Blatnik7349014}.
 
However, the operation of a detector in a so-called \textit{low ion backflow (low-IBF)} mode requires the application of HV settings which violate the rules of stable operation discussed above. This mode is envisaged for cases where the number of ions drifting back from the amplifications stage towards the drift volume needs to be maximally reduced. This is the case for Time Projection Chambers, where the back drifting ions build up space-charge, distorting the drift field, or photo-detectors, where the quality of a photo-sensitive cathode deteriorates upon bombardment of positive ions. In the low-IBF mode, the order of amplification within the stack is reversed and the fields are substantially altered with respect to the standard configuration (see~\cite{Adolfsson_2021, Ball_2014}, for example). 

In~\cite{Gasik:20198e} we show that these modifications result in a noticeable decrease in detector stability, which underlines the strong dependence of the latter on the HV settings. Discharge probability measured with the low-IBF configuration exceeds values obtained for the standard configuration by more than three orders of magnitude. The main reason for the stability deterioration in the low-IBF mode is the strong amplification in the last foil which, in combination with the pre-amplification of charges in the upper GEM foils, results in higher local charge densities in the last foil compared to those present in the standard configuration. In addition, the peculiar HV settings necessary to further minimize IBF require transfer fields above the last GEM in the amplification stack. This in addition alters the electric field configuration and the electron transport properties and may focus the charges into a lower number of individual GEM holes. 

The stability of a GEM stack operated in low-IBF mode can be, however, restored by adding a fourth GEM. In~\cite{Gasik:20198e} we show that the discharge probability measured with a quadruple GEM system improves significantly and the quadruple GEM configurations are more stable than a triple GEM operated in the low-IBW mode. It also shows that tuning the HV settings to obtain the lowest possible value of the IBF seriously affects the stability of the system.

\subsubsection{Multi-Micromegas}
\label{sec:stack:mmg}

There is not much experience with the multi-Micromegas stacks to date. First attempts performed in Saclay~\cite{JEANNEAU201094} focused on the ion backflow optimization showing the great potential of a double mesh configuration operated with highly optimized HV settings. Similar studies were launched recently at the University of Science and Technology of China, where a low ion backflow figure of \SI{<0.1}{\percent} at the gain of $\sim$2000 has been achieved with double Micromegas setups making them an attractive candidate for applications in gaseous photomultiplier tubes, RICH photoelectric readout, or high-rate TPCs~\cite{QI2020164282, ZHANG201878}. The results, however, were obtained with small 25$\times$25\,mm$^2$ prototypes and the feasibility of building large-area, uniform structures, is still to be presented.

In addition, the first attempt to quantify the discharge probability of a double Micromegas detector, based on 500\,LPI meshes of \SI{\sim40}{\percent} transparency each, is reported in~\cite{QI2020164282}. Preliminary results were obtained with an 8\,keV x-ray source 
For a direct comparison with the available literature on MPGD stability, dedicated measurements with alpha particles and hadron beams shall be considered. The possibility of employing meshes with different optical transparency in double Micromegas should also be studied. Although it will require a thorough optimization of the energy resolution and ion-stopping capabilities, the employment of low-transparency meshes may strongly improve the suitability of a Micromegas stack for operation in a harsh radiation environment. A scenario where a high-transparency mesh is used on top of the stack for high collection efficiency of primary electrons, and a low-transparency mesh is installed at the bottom for improved charge sharing between Micromegas cells, can be considered as a starting point for such an optimization.

\subsubsection{Hybrid stacks}
\label{sec:stack:hybrid}

Another category of stacks are hybrids of different MPGD technologies, with a GEM-like structure playing a role of the pre-amplification stage for the main amplification structure, usually a Micromegas, mounted below. The goal is to further suppress the ion backflow fraction and improve the discharge stability of the Micromegas. The former is reported to reach values down to \SI{0.1}{\percent}~\cite{CERN-LHCC-2015-002, AIOLA2016149, ihep_2018}. Introducing one or more GEM foils allows the operation of the Micromegas at a lower gain and enforces the spread of the charge cloud lowering the local charge densities arriving at the Micromegas stage. Discharge probability of the latter can be therefore substantially decreased by a factor of \SIrange[range-units=single]{10}{100}{} which has been confirmed with several measurements and simulations~\cite{MORENO2011135, Kane:2002gf, Procureur_2012}. The hybrid structures are successfully employed in a number of experiments, for example in the recent COMPASS RICH-1 upgrade~\cite{AGARWALA2018158, Agarwala:2018bku}, where THGEM\,+\,Micromegas structures are operated at high gains of $\sim$3$\times$10$^4$ with moderate and acceptable spark rates~\cite{fulvio:TUM}.  

However, in the case of a very demanding low-IBF operation, the stability of a hybrid stack is again compromised, as the gain of the Micromegas needs to be further increased. In the studies reported in~\cite{CERN-LHCC-2015-002} a 2GEM\,+\,MMG hybrid is compared to a quadruple GEM structure in the same conditions in a test-beam campaign at the CERN SPS fixed target facility. Full-size prototypes of both types were exposed to hadron showers from a thick iron target irradiated with a high-intensity secondary pion beam. Discharge rates measured with the 2GEM\,+\,MMG detector are three orders of magnitude higher than those recorded with the quadruple GEM solution. The results show the necessity for further investigation of the hybrid detectors in terms of their performance and stability optimization. Following the discussion above, low-transparency meshes could be combined with standard or small-pitch GEMs, keeping the energy resolution at, presumably, a satisfactory level while reducing the discharge probability of the Micromegas. This or similar configurations should be considered for further optimization in view of their application in future TPCs, for example at the planned Circular Electron Positron Collider~\cite{CEPCStudyGroup:2018ghi}.

\subsection{HV circuit optimisation}
\label{sec:optimum:hv}

Several recommendations for the safe operation of MPGD-based detectors have been worked out already in the early phase of their application (see for example~\cite{ALTUNBAS2002177, THERS2001133, Bozzo1462231}). These include protection resistors of $\mathcal{O}(1-10\,\mathrm{M\Omega})$, connected in series between the MPGD electrodes and their HV sources. 
In GEMs, where top electrodes are usually divided into $\sim$100\,cm$^2$ segments, each segment is protected individually. Although a spark discharge can develop, independently on the protection resistor, when the critical charge limit is reached, protection resistors decouple the HV voltage circuitry from the detector and quench the development of self-sustained discharges. In case of high ionization rates, currents induced on GEM electrodes will cause a voltage drop across the protection resistors which in turn will cause the gain drop and subsequently damp the possible discharge development. This, of course, needs to be balanced with the high-rate capability requirements of the detector (see also discussion on the resistive layers in \secref{sec:quench:resistive} below).
A similar effect can be achieved with Micromegas detectors where, apart from a moderate decoupling resistor connected to the non-segmented mesh, resistors of $\mathcal{O}(10)$\,M$\Omega$ can be connected to the individual readout strips causing only local voltage drops and reducing part of the active area being affected in case of a discharge~\cite{THERS2001133}. It should be noted, however, that Micromegas-based detectors evolved considerably in recent years and new techniques are being used to optimize their efficiency at high-ionization rates (see also \secref{sec:quench:resistive} and \secref{sec:quench:design}).

In addition to discharge quenching capabilities, a protection resistor reduces current flow through the structure in case of an electrical short development. In case the MPGD electrodes are divided into segments, each biased via its own protection resistor, a short in one or more segments can be tolerated without affecting the healthy part of the structure. The choice of resistor values is usually a compromise between gain fluctuations, caused by the voltage drop across the protection resistors, and hardware limitations to supply excess currents in case of a short in one or several MPGD segments. The effect of current flowing through a short on the overall potential difference across unaffected segments can be compensated by regulating the voltage source accordingly (see below).

The proper choice of resistive elements may also effectively quench the development of the so-called secondary discharges, developing in a gap between two GEMs or between a GEM and the readout electrode~\cite{DEISTING2019168}, not discussed in detail in this manuscript. Studies revealed that careful optimization of protection resistors on both sides of a GEM and reduction of parasitic capacitance, parallel to the transfer and induction gaps, shall be seriously considered while designing a multi-layer GEM detector~\cite{DEISTING2019168, Lautner_2019}. A list of clear recommendations for the HV scheme optimization can be found in~\cite{Lautner_2019}.

\subsection{Resistive layers}
\label{sec:quench:resistive}

Resistive electrodes are commonly employed in all types of MPGD-based detectors to improve spatial resolution and reduce the number of readout channels thanks to the charge-sharing capabilities but also as a method to provide spark protection to electronics and to reduce discharge probability. The recent review of the resistive gaseous detector can be found in~\cite{peskov_fonte_abbrescia}. Employed in an MPGD as an HV electrode, or a resistive layer capacitively coupled to the readout electrode (strips, pads), it creates a self-quenching mechanism for discharge development. It forces field reduction, by delaying charge evacuation, which results in reducing the discharge current, and thus discharge in the gas cannot be sustained. Moreover, the field reduction effects occur locally, in the region of the avalanche/streamer, and do not affect the entire detector area. The locality of the effect can be enhanced by segmented resistive electrodes.

With the gas discharge quenching capabilities, on the other hand, gain reduction comes into play at higher rates. In one of the pioneering studies with resistive Micromegas~\cite{ALEXOPOULOS2011110, BILEVYCH201166} it was shown that the signal amplitude slowly decreases with increasing rate, reaching a drop of \SI{\sim20}{\percent} at the rate of \SI{10}{\kilo\hertz\per\centi\meter\squared}. Similar rate capabilities were reported in the first tests of THGEMs with resistive electrodes~\cite{OLIVEIRA2007362}.

Further improvements in resistive-electrode technology continue. A diamond-like carbon (DLC), a class of meta-stable amorphous carbon material that contains both diamond-structure and graphite-structure~\cite{ROBERTSON2002129}, has recently attracted much attention in the MPGD community and is exploited to make resistive electrodes to suppress discharges occurring in MPGDs~\cite{OCHI_DLC,LV2020162759}). A promising development in this sector is represented by the \textmu-RWELL structures~\cite{Bencivenni_2015}, a single-sided GEM coupled to the readout anode through the material of high surface resistivity. A clear improvement of the stability in heavily irradiated environments was presented by comparing discharge probability values obtained with \textmu-RWELL and a standard 3-GEM setup, where both detectors showed similar performance~\cite{Bencivenni_2019}. In addition, it was shown that high-rate capabilities of the resistive structure can be restored by the proper grounding of the DLC layers assuring improved charge evacuation. The obtained performance figures together with simple construction (easy technology transfer to industry) make \textmu-RWELL a very attractive option for the application in, for example,  high-rate, large-area trackers.

The DLC coating can also reduce gain variations versus time and minimize charging-up effects, as shown in a recent study with standard THGEM foil, produced by an industrial print-circuit board fabrication procedure, coated with DLC films using magnetron sputtering technique~\cite{SONG2020163868}.

\subsection{Detector design optimisation}
\label{sec:quench:design}

Specific know-how was developed within the gaseous detector community providing a number of good practices to follow while designing a new detector. It is based on many years of experience gained by the MPGD developers and experimental groups using them. It is not possible to collect all recommendations in a single chapter, as many of the design considerations are also detector- and experiment-specific. Still, an approach to compile a list of the most necessary design considerations can be made, based on the available literature. The following overview focuses on GEM-like detectors but in most of the cases, the discussed recommendations can be applied to any kind of an MPGD.

\subsubsection{Segmentation of electrodes}

The most common design rule first proposed in~\cite{BACHMANN2002294} and applied thereafter to most of the GEM-based detectors is to subdivide their top electrodes into individual HV segments of \SI{\sim100}{\centi\meter\squared} area each. This way one can reduce the energy stored in a single-segment capacitor and limit the amount of charge released in case of an electrical discharge. Lower energy liberated in a discharge reduces the possibility of burning GEM foil constituents and developing an ohmic connection between two GEM electrodes. In case a segment does develop a short, the affected area of the detector is minimized, and the rest of the detector can be operated normally. 
Similar design can be introduced in other MPGD types (THGEMs, \textmu-RWELL, \etc) where electrodes can be conveniently segmented. In the case of Micromegas, where a standard mesh electrode cannot be easily divided (unless produced in a microbulk technology~\cite{DIAKAKI201846}), segmentation of the readout anode into strips or pads and their electrical connection play a similar role as individual GEM segments. In work~\cite{jonaphd} a so-called floating-strip Micromegas is proposed to minimize the effect of electrical discharges on the operation of a Micromegas detector. By grounding the mesh and applying HV to the resistive strips, capacitively coupled to the copper readout strips, one can effectively minimize the affected area of a discharge and thus the dead-time of the detector which reaches large values in case full Micromegas area discharges to the readout anode as in the case of normally operated bulk Micromegas detectors. 

Segmentation of a GEM electrode reduces also the probability of secondary discharges, which scales with an energy of the primary discharge\cite{BACHMANN2002294}. Based on studies presented in~\cite{Lautner_2019} the following points should be taken into account in the GEM detector design, to minimize the probability of secondary discharge occurrence.
\begin{itemize}
\itemsep0em 
    \item Capacitance of the transfer/induction gaps shall be reduced by limiting the single detector area, segmenting the bottom electrode of GEM foils, or increasing the gap size.
    \item Transfer and induction gaps in the detector should be kept uniform to avoid electric field enhancement and thus increased secondary discharge probability or even gas amplification in the gap. Using support structures is recommended, when possible.
\end{itemize}

It should be noted that the configuration with a segmented bottom electrode was usually avoided in the GEM community due to insufficient results and understanding of the secondary discharge formation process. With the newest results, this option is now being considered for future large-scale projects such as the CMS muon spectrometer upgrade with new GE2/1 and ME0 triple-GEM stations~\cite{merlin_propa,9508058}. 

\subsubsection{Number of amplification cells}

Following the results discussed in \secref{sec:mpgddisch}, optimization of a (TH)GEM geometry by reducing the pitch between holes can significantly increase the stability of the structure. This is related to the number of primary charges entering a single hole (an amplification cell). As the primary charge density is considered a driving factor of discharge formation, the larger number of holes in a structure directly influences the discharge stability of the latter. Reducing the pitch can be a highly beneficial design option, on the other hand, it needs to be balanced with the production capabilities and quality of a small-pitch structure.

The same argumentation can be presumably considered for Micromegas detectors, as pointed out in \secref{sec:mpgddisch}. Although further studies are still necessary to quantify the effect, a correlation between the size of a mesh cell and discharge probability has been shown~\cite{Gasik_2023}. Thus, a single Micromegas cell can be considered an independent amplification unit such as a GEM hole. This way, by varying the optical transparency of a mesh, one can adjust the discharge probability to a tolerable level, while keeping the ion backflow and energy resolution performance as required by the experiment. Further studies of this effect may provide valuable input for the performance optimization of Micromegas stacks or hybrid structures (see also the discussion in \secref{sec:stack:mmg} and~\ref{sec:stack:hybrid}).

\subsubsection{Elimination of electric field hotspots}

Although the discharge formation in a high-radiation environment is driven by charge densities approaching amplification cells, the elimination of high electric fields can minimize the occurrence of spurious discharges affecting the stable operation of an MPGD structure. Further design optimization should follow general recommendations for any kind of a gaseous detector, namely: avoid electric field enhancement around electrode edges, avoid sharp tips and corners of metallic parts, and maximize the distance between adjacent electrodes operated at different potentials or any HV electrode and the detector ground (this, of course, needs to be balanced with the requirement on minimum dead area). These recommendations highly depend on achievable production quality and can change in the future together with the improvement of micropattern technology. In order to avoid high peak field values in the case of Micromegas, also fundamental mesh parameters need to be considered. The size of wires, their density, and bending as well as the type of mesh influence the electric field values. Studies presented in~\cite{Bhattacharya_2020} show the peak-to-average field ratio can be reduced from three in the case of woven meshes to two for calendared ones. 

\subsection{Production optimisation}

The most efficient way to assure long-term stability and integrity of MPGD-based detectors, and to avoid instabilities caused by production defects and contamination, or design and assembly flaws, is a proper quality assurance (QA) and quality control (QC) of the produced MPGDs. Similarly to the design recommendations discussed above, there is no pre-defined list of tests to be performed with newly produced GEMs, Micromegas, or other types of structures. To large extent, it is driven by the requirements of the experiment but also the experience of users. However, recent developments for the ALICE TPC~\cite{Adolfsson_2021}, CMS GE1/1 muon~\cite{ABBAS2022166716} and ATLAS NSW~\cite{ALLARD2022166143} upgrades allowed to collect a high-statistics sample of MPGD structures which were produced and quality controlled using dedicated protocols, thoroughly worked out in the course of the projects. 
\section{Conclusions}

The constantly increasing requirements of a new generation of experiments in nuclear and particle physics are driving factors for the development of new detectors. Among many innovative technologies, Micro Pattern Gaseous Detectors have become widely used in high-rate physics experiments and are foreseen at future facilities. The key parameter for the long-term operation of MPGDs is stability against electrical discharges. Despite several open questions on primary and secondary discharge development, various results available in the literature and their interpretation contribute to the list of recommendations and good practices to be considered while designing new experimental apparatus. The stability limits of MPGD structures can serve as a reference for the optimization of such detectors. Detailed studies of discharge mechanisms in MPGD detectors result in a number of mitigation strategies that have been shown to efficiently minimize the probability of discharge occurrence. Many of them have been successfully implemented in recent large-scale projects. It should be noted that for each detector design, and operational conditions, careful optimization should be carried out to take into account project-specific constraints. The constant development of gaseous detector technology is essential for the realization of cutting-edge detectors in future nuclear and particle physics experiments.

\section*{Acknowledgements}
The publication is funded by the Open Access Publishing Fund of GSI Helmholtzzentrum fuer Schwerionenforschung.
This work was partially supported by the Deutsche Forschungsgemeinschaft - Sachbeihilfe [DFG FA 898/5-1].



\bibliographystyle{./elsarticle-modified.bst}
\bibliography{./references.bib}
\end{document}